\documentclass[twocolumn,showpacs,amsmath,amssymb,prl]{revtex4}

\usepackage{graphicx}
\usepackage{dcolumn}
\usepackage{bm}
\usepackage{placeins} 

\begin{document}

\setcounter{topnumber}{4} 
\setcounter{bottomnumber}{4} 
\setcounter{totalnumber}{10} 
\renewcommand{\textfraction}{0.15} 
\renewcommand{\topfraction}{0.85} 
\renewcommand{\bottomfraction}{0.70} 
\renewcommand{\floatpagefraction}{0.66}

\preprint{APS/123-QED}

\title{Leaky Membrane Dynamics}

\author{Robert S. Shaw$^{1,4}$}
\email{rob@protolife.com}
\author{Norman H. Packard$^{1,2,3}$}%
\email{n@protolife.com}
\affiliation{
\ \\
\mbox{$^1$ProtoLife Inc., 57 Post St. \#513, San Francisco, CA 91104}\\
\mbox{$^2$European Center for Living Technology, Dorsoduro 3859, 30123 Venezia, IT}\\
\mbox{$^3$Santa Fe Institute, Santa Fe, NM}\\
\mbox{$^4$Department of Nonlinear Dynamics, }\\
\mbox{Max Planck Institute for Dynamics and Self-Organization, Bunsenstra\ss e 10, 37073 G\"ottingen, Germany}
}
\date{\today}

\begin{abstract}
A concentration difference of particles across a membrane perforated by pores will induce a diffusive flux.  If the diffusing objects are of the same length scale as the the pores, diffusion may not be simple, objects can move into the pore in a configuration that requires them to back up in order to continue forward.  A configuration that blocks flow through the pore may be statistically preferred,
an attracting metastable state of the system.  This effect is purely kinetic, and not dependent
on potentials, friction or dissipation.  We discuss several geometries which generate this effect, and introduce a heuristic model which captures the qualitative features.
\end{abstract}
\maketitle


Managing flows of ions and molecules across membranes is a central task of living systems. 
Cell membranes contain a wide variety of pores, which can sort, drive, and rectify fluxes
of  different particles in both directions.  There are many models for the controlled leaks, and one-way flows of particular ionic species, which are useful to a cell \cite{hille:01}.  The simplest leak
is passive diffusion through a pore, driven by a concentration gradient.  If we assume noninteracting point particles, the flow can be described by Fick's law.  The flux is linear in
the concentration gradient across the membrane, the diffusion coefficient is
independent of concentration.  However, in a biological context, diffusing objects are of finite size,
often closely packed, and often not small compared to a pore diameter.  These conditions can
produce diffusion with a highly nonlinear dependence on concentration, and lead the system
to states far from the linear Onsager regime of nonequilibrium thermodynamics. The goal of this paper is to describe a common dynamic arising in such cases, generated solely by kinetic constraints and the geometry of the pore and diffusing objects.


A clear departure from linear diffusion takes place when some object enters a pore, and due to size or geometrical orientation cannot pass all the way through.  If there are many such blockers on one side of a membrane, carried in a bath of smaller free-flowing particles, we can produce a rectifier.  Flow in one direction sweeps blockers into pores, interrupting further flow.  Flow in the other direction clears the pores, re-establishing free flow.  This mechanism was in fact found in the squid giant axon, by Armstrong and Binstock in 1965 \cite{armstrong:65}.

A tacit assumption of many subsequent membrane studies is that there must exist forces which hold blockers in place, so that they are not dislodged by thermal fluctuations. Among proposed mechanisms are electrostatic attraction, due to charges fixed in the pore walls \cite{hille:78}, and microscopic flaps and tethers, to mechanically hold a blocker \cite{armstrong1997commentary,armstrong1977inactivation}.
Here we argue that kinetic constraints alone can readily perform this function. A blocker must backtrack to reopen a pore, and as concentration is increased, more and more objects obstruct this motion.  Thus pore clearance can become statistically unlikely, requiring a series of favorable fluctuations \cite{Shaw2007geometry,Packard2004asymmetric}.



Figure 1 illustrates a number of situations where blocking can arise, diffusing objects may arrive at a configuration that blocks flow through the pore.  This is a consequence of simple geometry only, and can occur in models employing Hamiltonian or Monte Carlo dynamics,
in continuous space or on a lattice, with two species of diffusing object, or just one.


The common theme of all of these models is that flow through a pore enters a blocked configuration, a dead end in the configuration space. In order for the block to be removed, objects must back up a finite distance along the path they travelled.  If there is a high concentration of objects in and around a pore, this retracing may be unlikely, and the system will be trapped in a long-lived metastable blocked state.  This blocked macroscopic state then becomes a type of attractor, in the sense that most microscopic configurations lead to it.

\begin{figure*}
\includegraphics[scale=0.5]{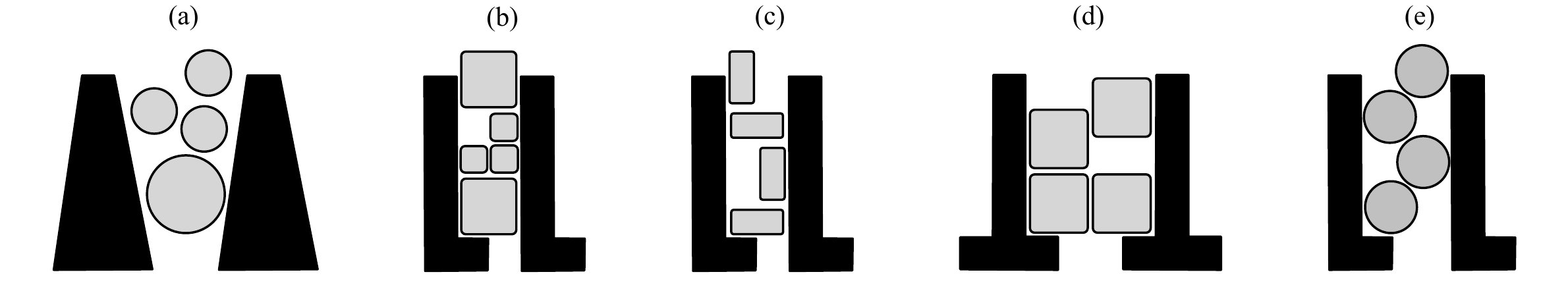}
\caption{\label{blocked}
Blocking of membrane pores with different pore geometries and different particle types. (a) Large and small disks moving in continuous space.  (b) Large and small squares moving on a lattice, with a spacing of the size of the small squares.  (c)  Dominos moving on a lattice; while in a pore, a domino cannot rotate.  (d)  Single species of square, moving on a lattice with spacing one-half the size of the square.  (e) Single species of disk, in the continuum.  In all cases, blockers must move upward to re-establish flow.  In model simulations, particles in (a) and (e) move in continuous two-dimensional space, governed by Hamiltonian dynamics.  Particles in (b-d) move on a lattice, with Monte Carlo updates.
}
\end{figure*}



Given a high enough concentration of particles, a pore becomes blocked, and remains blocked for a time that depends on concentration and pore geometry, which determine the strength of the attractor. When the stability of the attractor is weak, local density fluctuations around the pore may cause fluctuations between blocked and unblocked pore states, with a consequent intermittent flux transmission through the pore.  A hard disk simulation of a membrane consisting of a single pore
with the geometry of Fig.~1(a) displays this behavior \cite{PatchClamp}, as shown in Fig.~2.  This flux intermittency is qualitatively similar to data obtained from patch clamp observations of current through a single cell membrane pore \cite{hamill1981improved}.

\begin{figure}[b]
\includegraphics[scale=0.47]{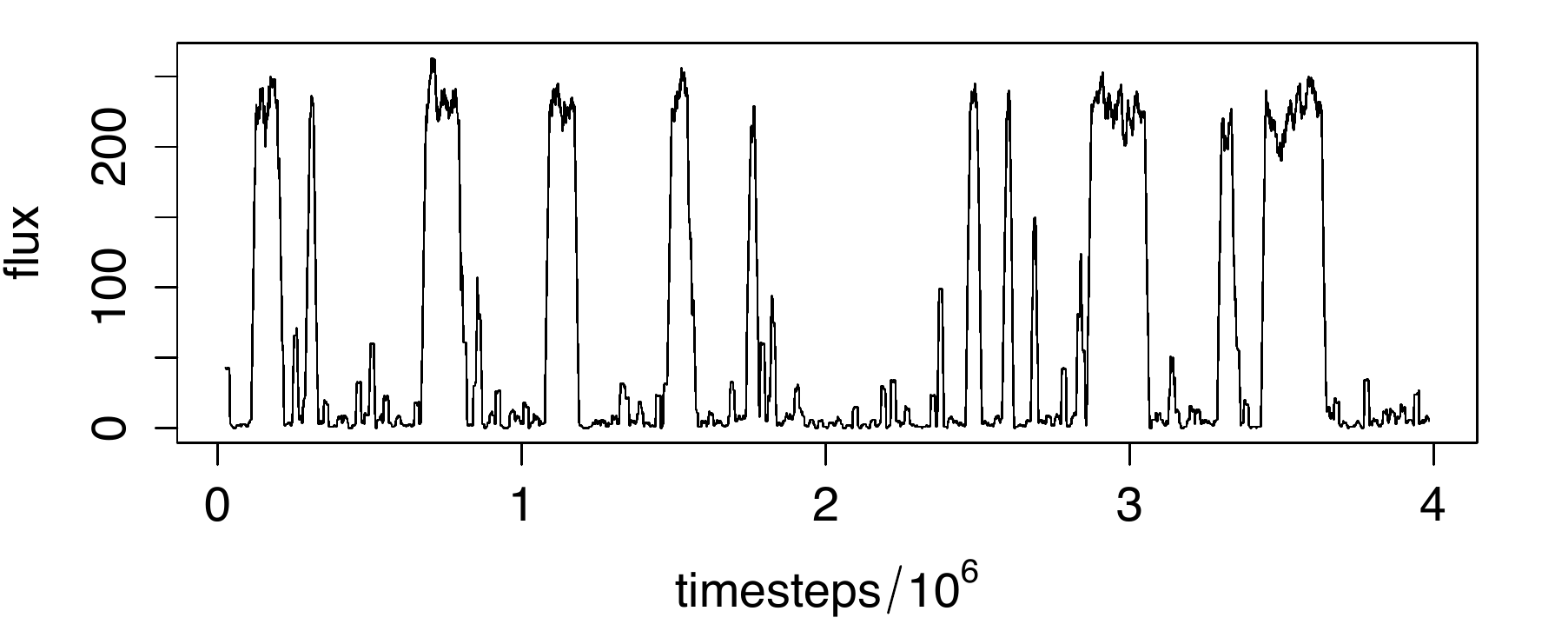}
\caption{\label{mem}
A "patch clamp" time series of the flux of small disks through a single pore  in the presence of blockers, as in the geometry of Fig.~1(a).  The particles are diffusing through the pore from a chamber kept at constant concentration, into a vacuum.  The flux values  are the number of particles passing through the pore in each interval of 1000 time steps \cite{PatchClamp}.
}
\end{figure}


The interactions between particles and pores that produce the blocking states illustrated in Fig.~1 are simple enough that we might expect this situation to occur quite generally.  We now present an illustrative model that uses macroscopic variables to characterize the blocking state of the whole membrane, capturing qualitative behavior common to all the various blocking scenarios.

The model membrane contains a large number of pores, each of which can either be open or closed.  For open pores, we assume Fickian diffusion and, for simplicity, that one side of the membrane is held at zero concentration.  The flux across the membrane is thus $J = fDC$, where $f$ is the fraction of open pores, $D$ is a diffusion coefficient, and $C$ the concentration on the occupied side of the membrane.
We describe the probabilities of a single pore entering or leaving the blocked state.

The probability that during a small time interval a blocker enters a pore we take to be proportional to concentration, and a coefficient reflecting the fraction of blockers, or blocking configurations,  $P_{\rm closes} = B C$.  The simulations indicate this assumption to be reasonable.  

 For a blocked pore to become open, 
 the blocker must experience a fluctuation which backs it out of the pore.  This will become increasingly unlikely as the concentration behind it increases, and as the pore length increases.  
 If we assume that a series of independent favorable fluctuations are required for the blocker to move backward a finite distance, individual fluctuation probabilities multiply, and we are led in the simplest case to an exponential, $P_{\rm opens} = e^{-GC}$, where $C$ is the concentration difference.  $G$ is a geometrical factor that
encodes the interaction between the diffusing blocker and the pore, and  would depend, for example, on the length of the pore, or the distance the blocker has to retrace.

Now we consider a chamber of unit volume, leaking through a membrane with many pores, into a vacuum,
and assume the number of pores is large enough so that we can treat the fraction of open pores, $f$, as a continuous variable.  We obtain a pair of coupled nonlinear first-order ODE's.
\[
{{dC}\over{dt}} = -fDC, \ \ \ \ \ \ \ \ \ \ \ \ \ \ \ 
{{df}\over{dt}} = (1-f)e^{-GC} - fBC
\]

The first describes the change of concentration in the chamber due to flux through the available open pores, and the second the change in the fraction of open pores due to the dynamics as described above. 

If the concentration in the chamber is held constant, we can solve for the flow as a function of concentration in the steady state, obtaining
\(
J(C) = {DC}/({BCe^{GC}+1}).
\)


Figure 3 displays the results of a hard-disk effusion simulation \cite{DumpHDisk}, along with curves from the above expression \cite{rectifyHDisk}.  The dotted curves are obtained from a Hamiltonian system of two sizes of elastic
hard disks moving in a chamber, one wall of which is a membrane perforated with tapered pores with the geometry of Fig. 1(a).  The smaller disks are able to effuse through the pores into a vacuum, any that do so are replaced in a random position in the chamber, thus keeping concentrations constant.  We examine
four different pore lengths. At a given pore, the outward flow may resemble the intermittent behavior of Fig.~2, but we take an average over multiple pores, and over a long time.

\begin{figure}[h]
\includegraphics[scale=.4]{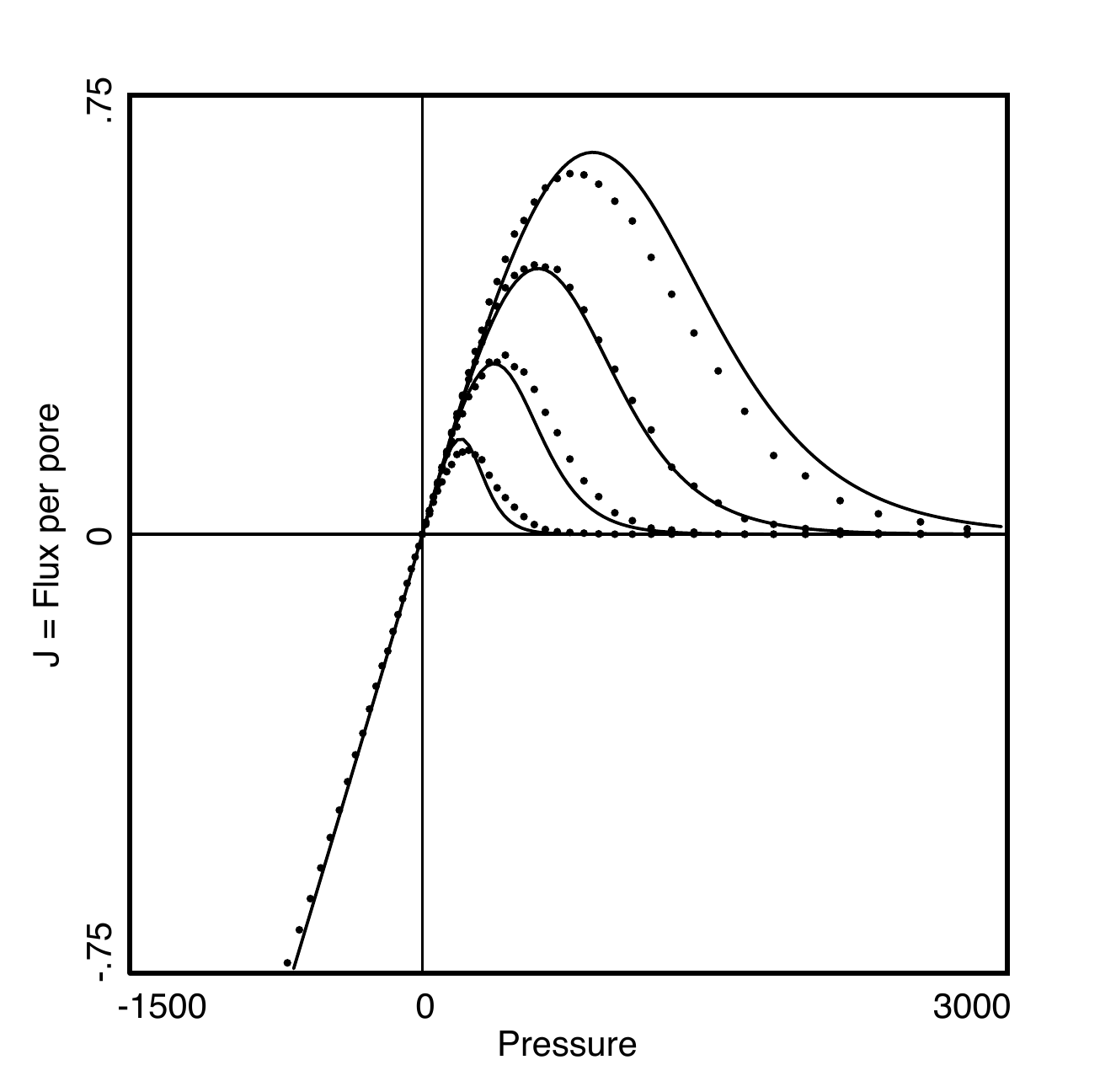}
\caption{\label{rec}
Steady-state flow of small disks through a membrane pore with the geometry of Fig.~\ref{blocked}(a), as a function of pressure in the chamber.  The dots are data points from a simulation, and the curves are generated by the steady-state equation above. The relative pore lengths are 2, 3, 5, and 10 (from the top curve down).  Flows through the reversed pore geometry are plotted as negative values.
}
\end{figure}

Plotted is the flux through the membrane as a function of the incremental pressure of small disks
added to the chamber, with a fixed number of large blocker disks present.
This pressure is the appropriate variable governing the dynamical flux in this Hamiltonian system,
and is computed from the concentration (disk area) using the hard disk equation of state \cite{Santos:1995p4001}.

For low concentrations of small disks, the flow is of Fick's law form, with an effective diffusion constant D.  But as the concentration is increased, large disk blockers are swept into pores,
and the mean flow is decreased.  At high concentrations, the probability that a blocker can back out of the pore becomes small, and the flow is reduced to near zero.

If the membrane geometry is reversed, that is, disks strike the membrane from the flat side, or bottom-up in Fig. 1(a), the large disks are not trapped in the pore by trailing small disks.  In this case, the flux remains a linear function of pressure up to high concentrations, these values are plotted as negative values in Fig.~3.  Note that the asymmetric geometry of the membrane leads directly to asymmetric diffusion in the two flow directions.

The solid curves of Fig.~3 are plotted from the analytic expression, we take G values in proportion to the simulation pore lengths, and fit the theoretical curves to the simulation data with a single overall scaling. The blocking factor B varies from 1, when all disks present in the chamber are large-disk blockers, to smaller values, as an increasing fraction of small disks are added \cite{rectifyHDisk}.
The diffusion constant D is found from the simulation data, in the linear low concentration range.

The fit is quite satisfactory, considering the pores in the simulation are actually conical, and that
density fluctuations of the hard disk gas are nontrivial \cite{Salacuse:2007p4002}.  In fact it will be more difficult for a blocker to escape than our exponential estimate predicts, because a fluctuation which empties a pore so that a blocker can back out will create a locally larger concentration near the mouth of the pore.  For extended particles, such a fluctuation will be reflected in a very large local pressure increase, and will be highly unlikely as concentration increases \cite{Santos:1995p4001,Salacuse:2007p4002}.  The model equations will not be
particularly accurate for all pore and blocker geometries.  But they may be a minimal set, describing deviations from the linear regime of nonequilibrium flow. All of the geometries in Fig.~\ref{blocked} generate qualitatively similar simulation curves.


This model has the form of a biased random walk, on a line 
with a reflecting barrier on one end (corresponding to the blocked end of the pore) and an absorbing barrier on the other end (the open end of the pore where the particle is absorbed by the bulk).  The bias is due to the mean flow.  Previous discussions of diffusion through pores have used similar, though unbiased, Brownian walk models.  However the physical interpretations of the random walk are completely different, for example describing a blocker tethered outside a pore \cite{millhauser1988diffusion}, or conformational transformations of the pore itself \cite{Goychuk:2002p1912}.


In our simple model, blockers are biased into a pore by a concentration difference, that is, a chemical potential.  We would like to note that a model of identical form is produced by considering a single charged blocker moving in a pore, biased by an electrical potential maintained across the membrane.  This might be a minimal model for a voltage-controlled rectifier, an important functional component of many cell membranes \cite{hille:01}.  Other models of voltage control feature charged sensor levers, and other mechanical  degrees of freedom which must move to restrict ion flow.  The model we present here can produce voltage control with no additional degrees of freedom, just a rigid pore with appropriate geometry.

%

Finally we examine the behavior of a leaky chamber filled initially at a high concentration, and then allowed to decline in time.  In Fig.~4 we compare the model equations with a simulation of the domino
geometry of Fig.~1(c) \cite{dominoRules}, though any of the models of Fig.~1 produce qualitatively similar results.
While the concentration gradient across the membrane is large, the leak rate is small.  But eventually the seal loosens, and there is a relatively sudden dumping of the contents
of the chamber.


For the domino simulations, note the large effect of fluctuations.  Even in this simulation, with a chamber initially containing 35,000 dominos, and a membrane of length 4000 with 500 pores, the onset of the high flux region is quite variable.  In both the simulation and the model equations, there is visible a short transient to an attracting state where most pores are blocked, from an initial condition where all pores are open.

\begin{figure}[h]
\includegraphics[scale=0.6]{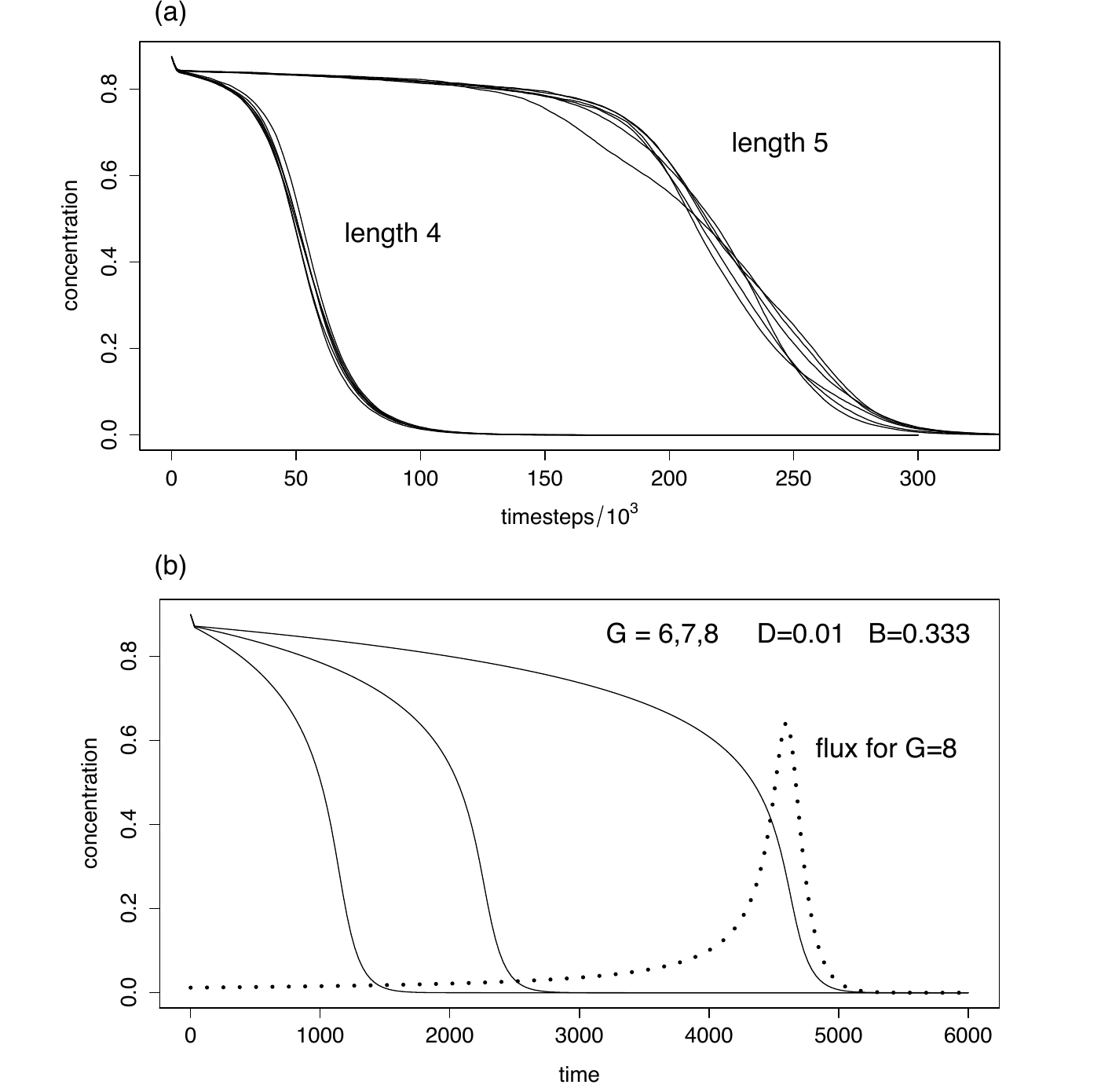}
\caption{\label{dump}
Simulation of the concentration of a leaky chamber of dominos, as a function of time.  Shown at the top are six runs with the same initial concentration, for two pore lengths with the geometry of  Fig.~1(c).  Below are curves generated by the model coupled equations, for a few values of G. The flux associated with the G=8 curve is plotted as a dotted line.
}
\end{figure}

Clearly, careful studies of the experimental consequences of this model are required.  For example,
the statistical properties of curves like that of Fig.~2 should be compared with real patch clamp data.
Another important issue is temperature dependence.  The escape probability of a blocker held by a kinetic constraint would not be expected to have the Arrhenius temperature dependence of a blocker held in a potential well.

Our discussion has emphasized the role of kinetic constraints and geometry in influencing flows of extended objects, via the simple requirement that a system back up a finite distance in its configuration space before proceeding further.  This picture of dead-end alleys in configuration space may be of some general use in modeling other nonequilibium systems.  However, geometry alone cannot be the whole story. A real membrane pore carries charged ions, and is constructed of proteins with charged residues; electric fields must play a central part.  A model including electric charges fixed to pore walls explains observed asymmetric diffusion in fabricated conical nanopores \cite{Siwy:2004p567}.
Also, living cell membranes may have evolved a number of intricate mechanisms for controlling ionic flow.
Nevertheless we  suggest that, due to its simplicity, the blocking mechanism described here will often be present, and moreover it may have been the dominant mechanism for simpler pores that occurred early in evolutionary history.


R. Shaw gratefully acknowledges the hospitality of Theo Geisel at the Max Planck Institute for Dynamics and Self-Organization, and financial support by the German Federal Ministry for 
Education and Research (BMBF) via the Bernstein Center for Computational 
Neuroscience (BCCN) G\"ottingen under Grant No. 01GQ0430.



\begin{thebibliography}{16}
\bibitem[{Pat()}]{PatchClamp}
\bibinfo{note}{A rectangular chamber of dimensions $2.0 \times 1.215$ has a
  single pore (geometry as in Fig. 1(a)) in the middle of one of the long edges of
  the rectangle, with pore length 0.85, pore width facing into the chamber of
  0.12, and pore width facing out of the chamber of 0.08. The chamber contained
  1200 disks of diameter 0.02, and a single disk of diameter 0.084 (the
  blocker). Any small particle that escapes is immediately put back into the
  chamber, randomly placed. Disks initially have unit velocity.}
\bibitem[{dom()}]{dominoRules}
\bibinfo{note}{Kinetic constraints for the dominos include the usual hard-core
  exclusion, but in addition may include constraints on domino rotations. The rule used in
  the present simulations is that a domino is allowed to rotate if both the
  present configuration and the future configuration have at least one end and
  one side free.}
\end{thebibliography}

\end{document}